\documentclass[twocolumn,showpacs,preprintnumbers,amsmath,amssymb]{revtex4-1}

\usepackage{afterpage}
\usepackage[dvipdfmx]{graphicx}
\usepackage{color,bm}

\begin{document}

\title{
Nondivergent deflection of light around a photon sphere of a compact object 
} 
\author{Ryuya Kudo}
\author{Hideki Asada} 
\affiliation{
Graduate School of Science and Technology, Hirosaki University,
Aomori 036-8561, Japan} 
\date{\today}

\begin{abstract} 
We demonstrate that the location of a stable photon sphere (PS) 
in a compact object is not always an edge 
such as the inner boundary of a black hole shadow, 
whereas the location of an unstable PS is known to be the shadow edge 
notably in the Schwarzschild black hole. 
If a static spherically symmetric (SSS) spacetime 
has the stable outermost PS, 
the spacetime 
cannot be 
asymptotically flat. 
A nondivergent deflection is caused for a photon traveling 
around a stable PS, 
though a logarithmic divergent behavior is known to appear 
in most of SSS compact objects with an unstable photon sphere. 
The reason for the nondivergence is 
that the closest approach of a photon 
is prohibited in the immediate vicinity of the stable PS 
when the photon is emitted from a source 
(or reaches a receiver) distant from a lens object. 
The finite gap size depends on the receiver and source distances 
from the lens as well as the lens parameters. 
The mild deflection angle of light can be approximated 
by an arcsine function. 
A class of SSS solutions in Weyl gravity 
exemplify the nondivergent deflection near the stable outer PS. 
\end{abstract}

\pacs{04.40.-b, 95.30.Sf, 98.62.Sb}

\maketitle

\section{Introduction}
Since the first measurement by Eddington and his collaborators 
\cite{Eddington}, 
the gravitational deflection of light has offered us 
a powerful tool for tests of gravitational theories including 
the theory of general relativity as well as 
for astronomical probes of dark matter. 
The Event Horizon Telescope (EHT) team has recently succeeded 
in taking a direct image of the immediate vicinity 
of the central black hole candidate of M87 galaxy \cite{EHT}. 
In addition, the same team has just reported 
measurements of linear polarizations 
around the same black hole candidate
\cite{EHT2021a}, 
which has led to an estimation  of the mass accretion rate 
\cite{EHT2021b}.
These observations have increased our renewed interest 
in the strong deflection of light in the strong gravity region. 

The strong deflection of light by Schwarzschild black hole 
was pointed out by Darwin \cite{Darwin}. 
This phenomena is closely related with 
a photon sphere (PS) that, with a horizon,  features 
black holes and other compact objects 
\cite{Perlick, CVE, Bozza2002, Hod2013, Sanchez, DFR, WLG, SYG, CGP, 
Tsukamoto2017, GW2016, Ohgami2015,Koga2016, Koga2018, Koga2019, Koga2020, Koga2021, Tsukamoto2020, Shiromizu, Yoshino2017,  Yoshino2020a, Yoshino2020b, Yang2020, Tsukamoto2021a, 
Izumi2021, Tsukamoto2021b, CG2016, Mishra2019, CG2021, Guo2021, Konoplya2021, Gan2021, Ghosh2021, Siino2021a, Siino2021b}. 
A photon surface for a less symmetric case 
is a generalization of PSs 
\cite{CVE}. 

Many years later, Bozza showed that 
the strong deflection behavior in 
Schwarzschild black hole 
can be well described 
as the logarithmic divergence in the deflection angle \cite{Bozza2002}. 
Such a strong deflection 
as the logarithmic divergence occurs 
also in other exotic objects such as wormholes. 
For instance, 
Tsukamoto conducted several extensions of Bozza method 
for the strong deflection of light 
\cite{Tsukamoto2017, Tsukamoto2020}, 
in which 
the logarithmic behavior is shown to be 
a quite general feature 
for a static and spherically symmetric (SSS) compact object  
that has a PS, 
e.g. Ellis wormholes. 
The logarithmic behavior in the strong deflection 
for a finite-distance receiver and source 
has been recently confirmed \cite{Takizawa2021}
by solving the exact gravitational lens 
that stands even for an asymptotically nonflat spacetime 
\cite{Bozza2008,Takizawa2020a,Takizawa2020b}.

There exists a single PS outside the horizon 
of the Schwarzschild black hole, 
while 
Ellis wormhole has a PS without horizons. 
For both cases, the number of PSs is one. 
Tsukamoto obtained the logarithmic behavior 
for such a SSS spacetime 
with a single PS, which was assumed to be unstable. 
In both of Schwarzschild  black hole and Ellis wormhole, 
there exists only the unstable PS. 
Cunha et al. have recently proven that ultra-compact objects 
have an even number of PSs, 
one of which is {\it stable}  
\cite{Cunha}. 
Regarding the interesting theorem, 
Hod has found an exception for horizonless spacetimes that 
possess no stable PS \cite{Hod2018}. 
What happens, 
if the PS is stable and a light ray is 
deflected around the stable PS? 

The main purpose of the present paper is to study 
 the deflection of light around a PS  
when the PS is stable 
 in a SSS spacetime. 
This situation has not 
often been considered in detail 
e.g. 
\cite{Darwin, Bozza2002, Tsukamoto2017, Tsukamoto2021a, Tsukamoto2021b}, 
except for Hasse and Perlick (2002) \cite{HP} 
that provided a theorem on a connection among three properties 
of (1) the presence of a PS 
for a saddle point case and a stable case as well as an unstable one, 
(2) the centrifugal force reversal 
and (3) infinitely many images 
in any SSS spacetime. 
Hasse and Perlick (2002), however, did not calculate 
the deflection angle of light, 
because they focused on the multiple imaging  
\cite{HP}. 
We shall show that, in stead of the logarithmic type of 
the strong deflection, 
a mild deflection is caused near the stable PS. 

This paper is organized as follows. 
In Section II, 
we reexamine the deflection angle of light 
when there exists a stable outer PS in a SSS spacetime. 
In Section III, we study how light is deflected 
in the presence of the stable outer PS. 
It is shown that the deflection behavior is not too strong 
to make the logarithmic divergence 
but mild enough to be approximated by an arcsine function. 
As an example for such a mild deflection of light, 
we consider a class of Weyl gravity model in Section IV. 
Section V concludes the present paper. 
Throughout this paper, we use the unit of $G = c = 1$.

\section{Deflection angle integral}
\subsection{SSS spacetime}
We consider a SSS spacetime, for which the metric reads 
\begin{equation}
  ds^2 = - A(r) dt^2 + B(r) dr^2 + C(r)(d \theta^2 + \sin^2 \theta d \phi^2) ,
 \label{metric}
\end{equation}
where $A(r) > 0$, $B(r) > 0$, and $C(r) > 0$ are assumed to be finite. 
If the SSS spacetime possesses a horizon, 
we focus on the outside of the horizon. 
We do not assume the asymptotic flatness of the spacetime. 
Actually, the present paper considers an asymptotically nonflat case. 
A lemma on this issue is given at the end of the present section.

\subsection{Photon orbits}
Without loss of generality, 
we can consider a photon orbit on the equatorial plane $(\theta = \pi/2)$ 
because of the spherical symmetry of the spacetime. 
On the equatorial plane in the SSS spacetime, 
a light ray has two constants of motion. 
One is the specific energy $E \equiv A(r) \dot{t}$ 
and the other is the specific angular momentum $L \equiv C(r) \dot{\phi}$, 
where the overdot denotes the derivative with respect to 
the affine parameter $\lambda$ along the light ray.  

By using the two constants $E$ and $L$, 
the impact parameter of light becomes 
$b = L/E$. 
Without loss of generality, we assume $b > 0$. 

In terms of $b$, 
the null condition $ds^2 = 0$ is rearranged as the orbit equation 
\begin{equation}
\dot r^2 + V(r) = 0 ,
\label{orbiteq}
\end{equation}
where $V(r)$ is defined as 
\begin{equation}
V(r) \equiv - \frac{L^2 F(r)}{B(r) C(r)} . 
\label{V}
\end{equation}
Here, 
\begin{equation}
F(r) \equiv \frac{C(r)}{A(r) b^2} - 1 .
\label{F0}
\end{equation}

The closest approach of a light ray is denoted as $r_0$, 
 which satisfies $V(r_0) = 0$ from the definition of $r_0$. 
Henceforth, evaluation at $r=r_0$ is indicated by 
the subscript $0$. 
For instance, $V(r_0) = 0$ is equivalent to $F_0 = 0$, 
where $F_0 \equiv F(r_0)$. 
By combining $F_0 = 0$ and Eq. (\ref{F0}), 
we obtain
\begin{equation}
b= \sqrt{\frac{C_0}{A_0}} . 
\label{impact}
\end{equation}
This offers a relation between $b$ and $r_0$.

Following Hasse and Perlick \cite{HP}, we 
use a particular form of the potential $\tilde V(r)$ 
that is defined as 
\begin{align}
\tilde{V}(r) \equiv \frac{A(r)}{C(r)} , 
\label{tildeV}
\end{align}
which is conformally invariant. 
Thereby Eq. (\ref{orbiteq}) is rewritten as 
\begin{equation}
A(r) B(r) \dot r^2 + L^2 \tilde{V}(r) - E^2 = 0 .
\label{tildeorbiteq}
\end{equation}
This simplifies the analysis of the photon sphere and its linear perturbation, 
because $V(r)$ depends on $b$ as well as $r$, 
whereas $\tilde{V}(r)$ does not include $b$. 
The potential $\tilde V(r)$ plays the central role 
in the proof of a theorem that clarifies a connection among three properties  
of the presence of a PS, 
the centrifugal force reversal 
and infinitely many images in any SSS spacetime 
\cite{HP}.

For the later convenience, 
we write down the first and second derivatives of $\tilde{V}(r)$, 
\begin{align}
\tilde{V}^{'}(r)
=& 
-\frac{A(r)D(r)}{C(r)} , 
\label{tildeV-1}
\\
\tilde{V}^{''}(r)
 =& 
- \frac{A(r)}{C(r)} \left[ D^{'}(r) - D(r)^2 \right] ,
 \label{tildeV-2} 
\end{align}
where the prime denotes the differentiation with respect to $r$, 
and the functions $D$  is defined as 
\begin{align}
D(r) 
&\equiv 
\frac{C'(r)}{C(r)} - \frac{A'(r)}{A(r)} .
\label{D1}
\end{align}

From Eq. (\ref{tildeorbiteq}), 
we obtain 
\begin{align}
[A(r) B(r)]^{'} \dot{r}^2 + 2 A(r) B(r) \ddot{r} + L^2 \tilde{V}^{'}(r) = 0 .
\label{ddotr}
\end{align}

Eqs. (\ref{tildeorbiteq}) and (\ref{ddotr}) tell that 
a photon orbit is a circle if and only if 
$r =  r_m$ satisfies 
$\tilde{V}(r_m) = (b_m)^{-2}$ and $\tilde{V}^{'}(r_m) = 0$, 
where $r_m$ denotes the radius of the PS 
and the subscript $m$ denotes the value at $r = r_m$. 
From $\tilde{V}(r_m) = (b_m)^{-2}$, 
the impact parameter for the PS 
is obtained as 
\begin{align}
b_m = \sqrt{\frac{C_m}{A_m}} , 
\label{bm}
\end{align}
where Eq. (\ref{tildeV}) is used.

\subsection{Classification of PS stability}
We consider a small displacement $\delta r$ around the PS orbit, 
$r = r_m + \delta r$. 
At the linear order in $\delta r$, Eq, (\ref{ddotr}) gives 
\begin{align}
 \frac{d^2}{d \lambda^2}(\delta r) 
 &= 
 - \frac{L^2 \tilde{V}^{''}(r_m)}{2A_mB_m}  \delta r 
 \notag\\
&= 
\frac{L^2 D^{'}_m}{2 B_m C_m} \delta r ,
\label{deltar}
\end{align}
where Eq. (\ref{tildeV-2}) and 
$D(r_m) = 0$ from $\tilde{V}^{'}(r_m) = 0$ 
are used.

The linear stability of the perturbed orbit is determined 
only by the sign of $D_m^{'}$, 
because 
$A(r) > 0$, $B(r) > 0$, $C(r) > 0$ and $L \neq 0$. 
A PS is stable if $D_m^{'} < 0$, 
whereas it is unstable for $D_m^{'} > 0$.

From Eq. (\ref{D1}), we obtain 
\begin{align}
D^{'}(r)
= \frac{C''(r)}{C(r)} - \frac{A''(r)}{A(r)} 
- D(r) \left(\frac{C'(r)}{C(r)} + \frac{A'(r)}{A(r)} \right) .
\label{D'}
\end{align}
On the PS, this is reduced to 
\begin{equation}
D'_m = \frac{C_m''}{C_m} - \frac{A_m''}{A_m} .
\label{D'm}
\end{equation}

Bozza and Tsukamoto assumed $D'_m > 0$, which 
means an unstable PS 
\cite{Bozza2002, Tsukamoto2017}. 
Henceforth, we focus on a stable case $D'_m < 0$. 
Such an unusual case is realized in a class of Weyl gravity model 
as shown in Section VI.

\subsection{Total angle integral}
From Eq. (\ref{orbiteq}), we obtain 
\begin{align}
\frac{d r}{d \phi} = \pm \sqrt{\frac{C(r)F(r)}{B(r)}} ,
\label{orbiteq-2}
\end{align}
for which we choose the plus sign without loss of generality. 
Integrating this from a source (denoted as S) to a receiver (denoted as R) 
leads to the total change in the longitudinal angle. 
\begin{align}
I_F(r_S, r_R, r_0)
&\equiv 
\sum_{i=S, R} 
\int_{r_0}^{r_i} \frac{dr}{\sqrt{\frac{C(r)F(r)}{B(r)}}} .
\label{IF-1}
\end{align}

Note that a conventional method discusses 
the total integral $I$ for the asymptotic receiver and source 
($r_R \to \infty$ and $r_S \to \infty$) 
e.g. \cite{Darwin, Bozza2002, Tsukamoto2017}, 
for which $I - \pi$ gives the deflection angle of light. 
On the other hand, the present paper considers finite distance 
between the receiver and source. 
In order to clarify this difference, we use $I_F$ in stead of $I$. 
Rigorously speaking, $I_F - \pi$ is not the deflection angle but 
$I_F$ is the dominant component of the deflection angle. 
This issue is beyond the scope of this paper. 
See e.g. References 
\cite{Ishihara2016, Ishihara2017, Ono2017, Ono2019U} 
on how to define geometrically the deflection angle 
for the finite-distance receiver and source.

Following Bozza and Tsukamoto 
\cite{Bozza2002, Tsukamoto2017}, 
we introduce a variable as 
\begin{equation}
z \equiv 1 - \frac{r_0}{r} ,
\label{z}
\end{equation}
to rewrite Eq. (\ref{IF-1}) as 
\begin{equation}
  I_F(z_S, z_R, r_0) = \sum_{i=S,R} \int_0^{z_i} f(z,r_0) dz , 
 \label{IF-2}
\end{equation}
where we define 
$z_i \equiv 1 - r_0/r_i$ $(i = R, S)$, 
\begin{align}
H(z, r_0) & \equiv \frac{CF}{B}(1-z)^4 ,
\label{H} 
\end{align}
and 
\begin{align}
f(z, r_0) & \equiv \frac{r_0}{\sqrt{H(z,r_0)}} .
\label{f}
\end{align} 

In the next subsection, 
we shall examine the integrand in Eq. (\ref{IF-2}).

\subsection{Analysis in the vicinity of PS}
By noting $F_0 = 0$, the function $H$ is expanded around 
$r = r_0$ ($z = 0$) as 
\begin{align}
H(z, r_0) = \sum_{n=1}^{\infty} 
c_n(r_0) z^n  ,
\label{H-2} 
\end{align}
where 
\begin{align}
c_1(r_0) 
&= \frac{C_0D_0r_0}{B_0} ,
\label{c1}
\\
c_2(r_0)
&= 
\frac{C_0}{B_0} \left[ D_0r_0\left( \frac{C'_0}{C_0} - \frac{B'_0}{B_0}
  -3 \right) + \frac{r_0^2}{2}(D_0^2 + D'_0) \right] ,  
  \label{c2}
\\  
c_3(r_0)
&= \frac{C_0}{B_0} \left[
  D_0r_0 \left( \frac{3B'_0r_0}{B_0} - \frac{3C'_0r_0}{C_0}
  - \frac{B'_0C'_0r_0^2}{B_0C_0} - 3 \right) \right. \nonumber \\
  &~~~~~~~~~ + \frac{r_0^2}{2}(D_0^2 + D'_0) \left( \frac{C'_0r_0}{C_0} -
  \frac{B'_0r_0}{B_0} -2 \right) \nonumber \\
  &~~~~~~~~~ \left. + \frac{r_0^3}{6} \left( D_0^3 +3D'_0D_0 + D''_0 \right) \right] .
 \label{c3}
\end{align}

When the closest approach is located near the PS ($r_0 = r_m$), 
Eqs. (\ref{c1}) - (\ref{c3}) become 
\begin{align}
c_1(r_m) 
&= 0 , 
\label{c1-PS}
\\
c_2(r_m) 
&= \frac{r_m^2 C_m D_m'}{2 B_m} ,
\label{c2-PS}
\\
c_3(r_m) 
&= \frac{r_m^2 C_m}{2 B_m} 
\left[ D_m' \left( \frac{C_m' r_m}{C_m} - \frac{B_m' r_m}{B_m} -2 \right) 
+ \frac{r_m D_m''}{3} \right] . 
\label{c3-PS}
\end{align}

From Eqs. (\ref{c1-PS}) - (\ref{c3-PS}), 
we find 
\begin{equation}
H(z,r_m) = c_2(r_m) z^2 + O(z^3) . 
\label{H_m}
\end{equation}

If the PS is unstable (stable), 
namely $D^{'}_m > 0$ ($D^{'}_m < 0$), 
then, 
$c_2(r_m) > 0$ ($<0$). 
As pointed by Tsukamoto \cite{Tsukamoto2017},  
the angle integral $I_F$ by Eq. (\ref{IF-1}) 
is divergent logarithmically. 
On the other hand, 
the stable PS case ($c_2(r_m) < 0$) 
is investigated below in detail. 

Before closing this section, 
we mention a relation of 
the emergence of the 
stable outermost PS 
(SOPS) 
and the spacetime 
asymptotic flatness. 

{\bf \noindent Lemma}\\
If a SSS spacetime has the SOPS, 
the spacetime 
cannot be asymptotically flat, 
for which 
$\tilde{V}^{'}(r)= [A(r)/C(r)]^{'}$  
is positive everywhere outside the SOPS. 

{\bf \noindent Proof}\\
We denote the radius of the SOPS as $r_{SOPS}$. 
At the location of the SOPS,  
$\tilde{V}^{'}(r_{SOPS}) = 0$ 
and $\tilde{V}^{''}(r_{SOPS}) > 0$. 
There exist no PSs outside of the SOPS, 
because the SOPS is the outermost PS. 
Therefore,  $\tilde{V}^{''}(r) > 0$ for $r > r_{SOPS}$. 
This means that  $\tilde{V}^{'}(r)$ is 
an {\it increasing} function of $r$ 
when $r > r_{SOPS}$. 
Hence,  $\tilde{V}^{'}(r) > 0$ for $r > r_{SOPS}$. 

Here, we add an assumption that the spacetime were asymptotically flat. 
Then, we can employ a coordinate system 
in which Eq. (\ref{metric}) approaches 
the Minkowski metric in the polar coordinates 
asymptotically as $r \to \infty$. 
Namely, $A(r) \to 1 + O(1/r)$ and $C(r) \to r^2 +O(r)$, 
which lead to $A^{'}(r) \to O(1/r^2)$ and $C^{'}(r) \to 2r + O(r^0)$. 
Thereby, 
\begin{align}
\tilde{V}^{'}(r) 
&= \left(\frac{A(r)}{C(r)}\right)^{'} 
\notag\\
&= -\frac{2}{r^3} + O\left(\frac{1}{r^4}\right) 
\notag\\
&\to 0 ,
\label{Limit}
\end{align}
in the limit as $r \to \infty$. 
This means that 
$\tilde{V}^{'}(r)$ has an extremum between $r_{SOPS}$ and $r=\infty$, 
because of its continuity. 
This contradicts with that $\tilde{V}^{'}(r)$ is an increasing function 
for $r > r_{SOPS}$. 
Therefore, the spacetime is not asymptotically flat. 
A proof of the lemma is thus completed.

According to Reference \cite{Guo2021}, 
if an axisymmetric, stationary and asymptotically flat spacetime 
posses light rings (LRs), 
the outest LR  is unstable. 
This means that 
if a SSS spacetime with PSs is asymptotically flat, 
the outest PS is unstable. 
The contraposition of this is that, 
if the outest PS in a SSS spacetime with PSs is stable, 
the spacetime is not asymptotically flat. 
This proves a part of the above lemma but 
tells nothing about the positivity of $\tilde{V}^{'}(r)$ for $r > r_{SOPS}$.

On the other hand, 
it is clear that the asymptotic flatness is allowed, 
if the outermost PS (OPS) is unstable 
($\tilde{V}^{''}(r_{OPS}) < 0$). 
In the rest of this paper, we consider that a receiver and source 
are located outside of a stable outer PS. 
However, it is not specified below whether or not it is the outermost. 

Before closing the section, let us briefly mention the stability of 
a spacetime that admits a stable PS. 
Horizonless ultracompact objects with a stable PS 
may suffer from instabilities due to slowly (at most logarithmically) decaying 
of perturbations leading to the formation of a trapped surface 
\cite{Cardoso2014, Keir, Cunha, AMY}. 
In section VI, therefore, we consider a black hole model with 
a stable PS in Weyl gravity. 
On the other hand, 
the above lemma may suggest another possibility 
compatible with the instability arguments 
in References \cite{Cardoso2014, Keir, Cunha, AMY}. 
One such candidate is a compact object 
without a black hole horizon but with a stable PS and 
a cosmological horizon that is consistent 
with the asymptotic nonflatness. 
Its stability issue is beyond the scope of the present paper.

\section{Mild deflection near the stable outer PS}
\subsection{Stability classification of the outer PS}
In the neighborhood of the closest approach $z \sim 0$, 
higher order terms of $H(z,r_m)$ in Eq. (\ref{H-2}) are negligible 
compared with $z$ and $z^2$ terms. 
The dominant part of $f(z, r_0)$ for $r_0 \sim r_m$ thus becomes 
\begin{equation}
f_D(z, r_0) \equiv 
\frac{r_0}{\sqrt{h(z)}} ,
\label{fD}
\end{equation}
where $h(z)$ is defined as 
\begin{align}
h(z) = c_1(r_0)z + c_2(r_0)z^2 . 
\label{h}
\end{align}

Associated with $f_D(z, r_0)$, we define the dominant part of 
the total angle integral as 
\begin{equation}
 I_{FD}(z_R, z_S, r_0) 
 \equiv 
 \sum_{i=S, R}
 \int_0^{z_i} f_D(z,r_0) dz . 
\label{IFD}
\end{equation}

Before starting calculations of the angle integral, 
first we investigate a photon orbit in the stable PS case. 
If $c_1(r_0) < 0$, $h(z)$ is always negative for $z>0$. 
This is in contradiction with the nonnegativity of $h(z)$. 
Hence, the case of $c_1(r_0) < 0$ is discarded. 
If $c_1(r_0) = 0$, then, $h(z) = c_2(r_0) z^2$. 
The nonnegativity of $h(z)$ together with $c_2(r_0) < 0$ 
admits only $z=0$. 
This orbit is a circle. 
We do not discuss this special case any more. 
For the last case $c_1(r_0) > 0$, 
the nonnegativity of $h(z)$ provides a nontrivial situation; 
the allowed region for $z$ is 
\begin{align}
0 \leq z \leq -\frac{c_1(r_0)}{c_2(r_0)} . 
\end{align}

This means that only bound orbits are allowed, 
whereas the scattering orbits are prohibited. 
Note that $h(z) = 0$ does not make the integral in Eq. (\ref{IFD}) 
divergent. 
Indeed, $h(z) \sim c_1(r_0) z$ near a point $z = 0$ 
and 
$h(z) \sim -c_1(r_0) [z + c_1(r_0)/c_2(r_0)]$ 
near a point $z \sim -c_1(r_0)/c_2(r_0)$. 
At the two points, therefore, the integral of $h(z)$ is not divergent 
as $\sim \int z^{-1/2} dz \sim 2 \sqrt{z}$.  
The two points $z = 0$ and $z \sim -c_1(r_0)/c_2(r_0)$ 
are the periastron and apastron, respectively.

\subsection{Angle integral for the stable outer PS}
Henceforth, we focus on the case of $c_1(r_0) > 0$, 
for which 
the receiver and source positions satisfy 
\begin{equation}
0 < z_i \leq -\frac{c_1(r_0)}{c_2(r_0)} . 
\label{allowed}
\end{equation}
Eq. (\ref{IFD}) is integrated as 
\begin{align}
I_{FD}(z_R, z_S, r_0) 
=& 
\frac{r_0}{\sqrt{- c_2(r_0)}} 
\notag\\
&
\times 
\left[\pi - \sum_{i = S,R} \arcsin 
\left( 1 + \frac{2 c_2(r_0)}{c_1(r_0)} z_i \right) \right] .
\label{IFD-2} 
\end{align}

It follows that the arcsine term in Eq. (\ref{IFD-2}) 
is well defined for the allowed region 
by Eq. (\ref{allowed}), 
because 
\begin{equation}
  -1 \leq 1 + \frac{2 c_2(r_0)}{c_1(r_0)} z_i < 1 .
\label{arcsin}
\end{equation}

\subsection{$I_F$ in terms of $b$}
In most of lens studies including References 
\cite{Bozza2002, Tsukamoto2017}, 
it is convenient to express the deflection angle 
in terms of the impact parameter in stead to $r_0$, 
mainly because 
$b \sim D_L \theta$ for the lens distance $D_L$ 
and the image angle direction $\theta$. 

Therefore, we look for an approximate expression of $b$. 
Near the PS, $c_1(r_0)$ and $b(r_0)$ are Taylor-expanded as 
 \begin{align}
c_1(r_0) 
&= 
\frac{C_m r_m D'_m}{B_m}(r_0 - r_m) + O((r_0 - r_m)^2) , 
\label{c1m}
\\
b(r_0) 
&= 
b_m + \frac{b_m D'_m}{4}(r_0 - r_m)^2 + O((r_0 - r_m)^3) . 
  \label{bm}
\end{align}

\begin{figure}
\includegraphics[width=8.6cm]{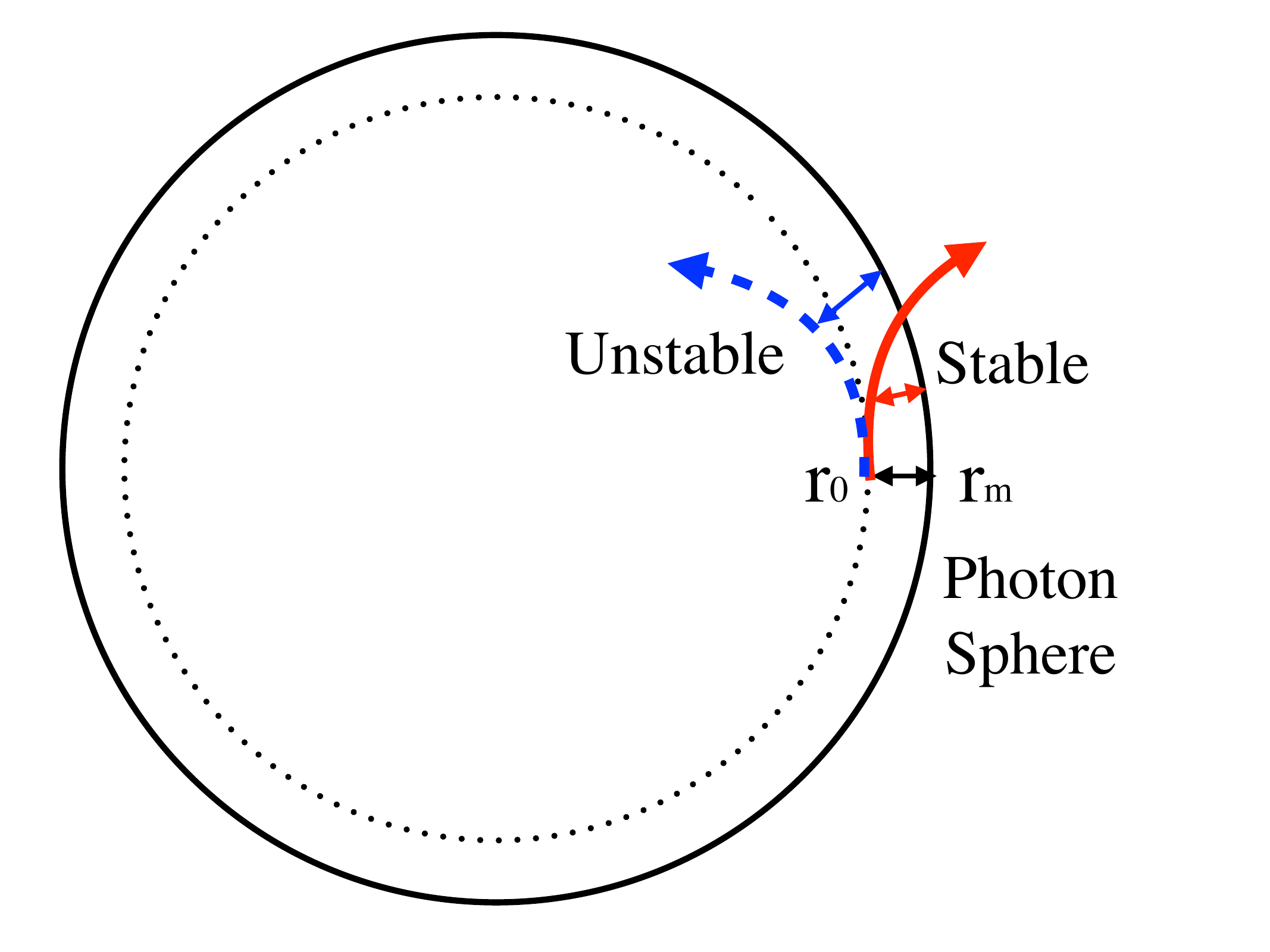}
\caption{
Schematic figure of photon orbits near a PS. 
The PS is denoted by a solid circle with radius $r_m$. 
The dotted circle has radius $r_0$, which is slightly 
smaller than $r_m$. 
Namely, $r_m - r_0$ is taken 
as a perturbation around the PS. 
The initial position of the photon is $r_0$, at which 
the initial radial velocity is vanishing. 
The thick dashed blue (in color) arrow denotes a photon motion 
when the PS is unstable. 
The orbital deviation grows because of being unstable. 
Therefore, it is difficult for the inner photon to escape to a far region. 
This is why the earlier papers focused on  $r_0 > r_m$ 
when an unstable PS is assumed 
\cite{Bozza2002, Tsukamoto2017}. 
On the other hand, 
the thick solid red (in color) arrow denotes a photon motion  
when the PS is stable. 
The orbital deviation gets smaller because of being stable. 
It is thus possible that the inner photon ($r_0 < r_m$) 
crosses the PS to reach a distant observer. 
} 
\label{fig-orbits}
\end{figure}

From Eq. (\ref{c1m}), we find 
\begin{align}
r_0 < r_m ,
\end{align}
where $c_1(r_0) > 0$ and $D^{'}_m < 0$ are used. 
This inequality means that the light ray passes by the slight inside of the PS. 
This is because the PS is stable. 
This unusual behavior of the photon orbit 
implies also 
\begin{align}
b < b_m ,
\end{align}
from Eq. (\ref{bm}). 
See Figure \ref{fig-orbits} for the photon orbit behavior near the PS.

By combining Eqs. (\ref{c1m}) and (\ref{bm}), 
we obtain, near the PS,  
an approximate relation between $c_1(r_0)$ and $b$ as 
\begin{align}
c_1(r_0) 
\approx 
\frac{2 r_m C_m \sqrt{-D'_m}}{B_m} \left( 1 - \frac{b}{b_m} \right)^{\frac{1}{2}} . 
\label{c1b}
\end{align}
By substituting this into Eq. (\ref{IFD-2}), 
we obtain 
\begin{align}
&I_{FD}(z_R, z_S, b) 
\notag\\
&\simeq 
\frac{r_m}{\sqrt{- c_2(r_m)}} 
\notag\\
&~
\times 
\left[\pi - \sum_{i = S,R} \arcsin 
\left\{ 1 - \frac{r_m \sqrt{-D'_m}}{2} \left( 1 - \frac{b}{b_m} \right)^{-\frac{1}{2}} z_i \right\} \right] .
\label{IFD-approx}
\end{align}

By direct calculations for a photon traveling near the PS, 
Eq. (\ref{allowed}) is rearranged as 
\begin{equation}
  0 < z_i \le \frac{4}{r_m \sqrt{-D'_m}} 
\left( 1 - \frac{b}{b_m} \right)^{1/2} , 
\label{allowed-approx}
\end{equation}
for which the arcsine function in Eq. (\ref{IFD-approx}) is well defined. 

In order to obtain 
Eq. (\ref{IFD-approx}) as an approximate estimation of $I_F$ 
in terms of $b$, 
we have used the Taylor series expansion as Eq. (\ref{bm}), 
which is valid if $b_m > b_m |D^{'}_m| (r_0 - r_m)^2 /4$. 
This requires that the closest approach of light and the PS  
are close enough to satisfy
\begin{align}
|r_0 - r_m| < \frac{2}{\sqrt{|D^{'}_m|}} .  
\end{align}
Yet, a lower bound on $|r_0 - r_m|$ 
exists in the neighborhood of the PS 
as shown below.

\subsection{Discontinuity between the closest approach and the stable PS}
Eq. (\ref{allowed-approx}) suggests a proposition on the existence of 
a gap between the allowed closest approach and the stable PS. 

{\bf \noindent 
Proposition}\\
In a SSS spacetime which possess a stable outer photon sphere, 
the closest approach of a photon from a source (or to a receiver) 
located at finite distance from the lens object 
is not allowed in the infinitesimal neighborhood of the stable PS.

{\bf \noindent
Proof}\\
This proposition can be proven by contradiction as follows. 
We consider a receiver (or a source) at finite distance from the lens, 
namely  $r_i > r_0$, 
which leads to $z_i > 0$. 
We assume the closest approach of a photon orbit were 
in the infinitesimal neighborhood of the stable PS. 

On the stable PS, 
$F_m = F^{'}_m = 0$, 
while $F^{''}_m < 0$. 
The last inequality comes from 
the unstable condition $D^{'}_m < 0$. 
The function $F$ near the PS is thus Taylor-expanded as 
\begin{align}
F_0 = \frac12 F^{''}_m (r_0 - r_m)^2 
+ O\left( (r_0 - r_m)^3 \right) . 
\label{F0-exp} 
\end{align}
 From Eq. (\ref{H}), 
 $H$ for $r_0 \approx r_m$ is expanded as 
 \begin{align}
 H = \frac{C_m F^{''}_m}{2 B_m}(1-z)^4 (r_0 - r_m)^2 
 + O\left( (r_0 - r_m)^3 \right) ,
 \label{H-exp2}
 \end{align}
 where Eq. (\ref{F0-exp}) is used. 
 
 By using $B_m > 0$, $C_m > 0$ and $F^{''}_m < 0$ 
 for Eq. (\ref{H-exp2}), 
we find $H < 0$ 
if $r_0 - r_m$ is sufficiently small. 
$H < 0$ contradicts with the existence of the photon orbit. 
Our proof is finished. 

The above proposition prohibits 
the closest approach in the infinitesimal neighborhood 
of the stable PS. 
However, it does not tell about a size of the gap 
between the allowed closest approach and the stable PS. 
In order to discuss the gap size, 
we use Eq. (\ref{allowed-approx}), 
which is rewritten as
\begin{align}
\frac{b}{b_m} 
\leq 
1 - \frac{r_m^2 z_i^2 (- D_m^{'})}{16} .
\label{b-inequality}
\end{align}
Note that Eq. (\ref{allowed-approx}) 
is based on a quadratic approximation up to $z^2$. 

For finite $z_i$ and $D_m^{'} < 0$, 
Eq. (\ref{b-inequality}) demonstrates that 
$b$ is not allowed in the infinitesimal neighborhood of $b_m$. 
From Eq. (\ref{b-inequality}),  
the upper bound on $b$ is 
\begin{align}
b_{max} 
= 
b_m - \frac{b_m r_m^2 z_L^2 (- D_m^{'})}{16} . 
\end{align}
The gap size $\Delta b \equiv b_m - b_{max}$ 
is thus given by 
\begin{align}
\Delta b 
= 
\frac{b_m r_m^2 z_L^2 (- D_m^{'})}{16} ,
\label{Deltab}
\end{align}
where $z_{L}$ is the larger one of $z_R$ and $z_S$, 
namely the value of $z$ for the more distant one 
of the receiver and source.

A separate treatment is needed for 
the marginal PSs ($D'_m = 0)$ 
e.g.  
\cite{CT,Tsukamoto2020}.

\subsection{The dominant part and the remainder}
Before closing this section, 
we shall confirm that $I_{FR} \equiv I_F - I_{FD}$ 
is really the remainder. 
It is written simply as 
\begin{equation}
I_{FR}(z_R, z_S, r_0) 
\equiv 
\sum_{i = R, S} 
\int_0^{z_i} f_R(z, r_0) dz , 
\label{IFR}
\end{equation}
where 
$f_R(z,r_0) \equiv f(z, r_0) - f_D(z, r_0)$. 

From Eq. (\ref{IFR}), we obtain, for $z_i < 1$, 
\begin{align}
I_{FR}(z_R, z_S, r_0) 
= \sum_{i = R, S} [f_R(z_i, r_0) z_i +O(z_i^2)] . 
\label{IFR-approx}
\end{align}

For $r_0 \sim r_m$, 
$f(z, r_m) \sim f_D(z, r_m)$ when $z < 1$. 
Hence, by using the Taylor-expansion method, 
one may think that $f_R(z, r_m) = f(z, r_m) - f_D(z, r_m) = O(z)$. 
However, let us more carefully 
examine an asymptotic expansion of $f_R(z, r_0)$ 
for $z < 1$ 
\cite{footnote-expansion}. 
By straightforward calculations, 
the asymptotic expansion of $f(z, r_0)$ for small $z$ 
is obtained as 
\begin{align}
f(z, r_0) 
=& 
\frac{r_0}{\sqrt{z c_1(r_0)}} 
- \frac{r_0 c_2(r_0)}{2 c_1(r_0)^{3/2}} \sqrt{z} 
+ O(z^{3/2}) . 
\label{f-exp}
\end{align}
Similarly, 
the asymptotic expansion of 
$f_D(z, r_0)$ is 
\begin{align}
f_D(z, r_0) 
=& 
\frac{r_0}{\sqrt{z c_1(r_0)}} 
- \frac{r_0 c_2(r_0)}{2 c_1(r_0)^{3/2}} \sqrt{z} 
+ O(z^{3/2}) . 
\label{fD-exp}
\end{align}
By bringing together Eqs. (\ref{f-exp}) and (\ref{fD-exp}), 
we find 
\begin{align}
f_R(z, r_0) = O(z^{3/2}) . 
\label{fR-exp} 
\end{align}

By using this for Eq. (\ref{IFR-approx}), we find 
\begin{align}
I_{FR}(z_R, z_S, r_0) = O(z_R^{5/2}, z_S^{5/2}) . 
\label{IFR-approx2}
\end{align}

On the other hand, Eq. (\ref{IFD-2}) is expanded, 
for small $z_R$ and $z_S$, as
\begin{align}
I_{FD}(z_R, z_S, r_0) 
& = 
\frac{2 r_0}{\sqrt{c_1(r_0)}} 
(\sqrt{z_R} + \sqrt{z_S}) 
+ O(z_R, z_S) ,
\label{IFD-approx2} 
\end{align}
where 
$\arcsin (1 -\varepsilon) = \pi/2 -\sqrt{2\varepsilon} + O(\varepsilon)$ 
is used.

For $z_R, z_S \ll 1$, Eqs. (\ref{IFR-approx2}) and (\ref{IFD-approx2}) 
lead to 
$I_{FR}(z_R, z_S, r_0) \ll I_{FD}(z_R, z_S, r_0)$. 
Therefore, we can safely say that 
$I_{FD}$ is the dominant part and 
$I_{FR}$ is the remainder.

\section{Example: A class of Weyl gravity model}
In the Weyl gravity model, Mannheim and Kazanas 
found a class of SSS solutions \cite{MK}. 
The metric reads
\begin{equation}
 ds^2 
 = - g(r) dt^2 + g(r)^{-1} dr^2 
 + r^2(d\theta^2 + \sin^2 \theta d\phi^2) ,
\label{ds-Weyl}
\end{equation}
where 
\begin{equation}
g(r) = 1 - 3 \beta \gamma 
- \frac{\beta (2 - 3 \beta \gamma)}{r} + \gamma r - \kappa r^2 , 
\end{equation}
and we consider $g(r) > 0$, namely the outside of the horizon. 

The allowed region for the existence of the stable outer PS  
is \cite{TH} 
\begin{align}
\beta > 0 , 
\\
\gamma < 0 , 
\\
\kappa < 0 . 
\end{align}
In this section, we focus on this parameter region.

\begin{figure}
\includegraphics[width=8.6cm]{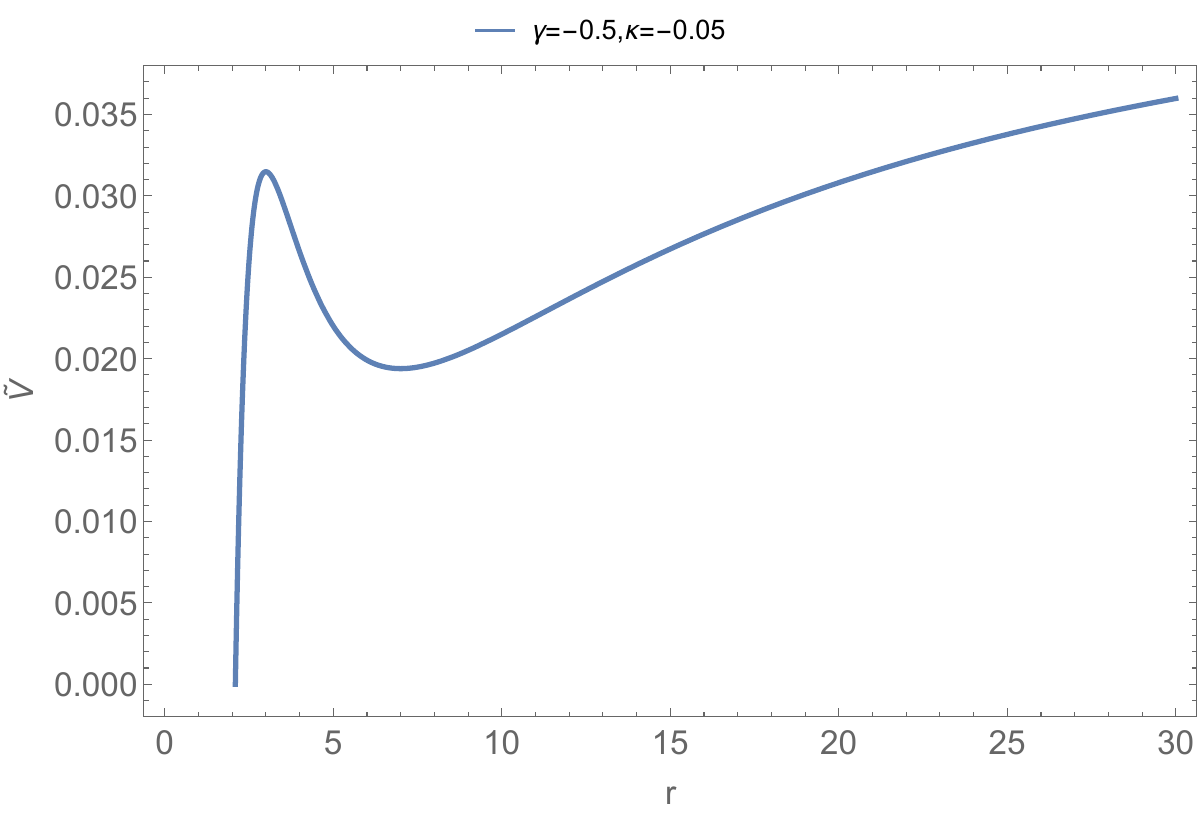}
\caption{
Effective potential $\tilde{V}(r)$ 
in the Weyl gravity model. 
The solid blue (in color) curve 
corresponds to $\gamma = -0.5$ and $\kappa = -0.05$ 
in the unit of $\beta =1$, 
which leads to the SOPS at $r_m = 7$. 
From $(b_m)^{-2} = \tilde{V}(r_m) = 0.0193878$, 
$b_m = 7.18184$ is obtained. 
$\tilde{V}(r)$ does not vanish for large $r$, 
because the spacetime is not asymptotically flat. 
} 
\label{fig-V}
\end{figure}

We solve $D(r) = 0$ to find two roots as 
$r = 3\beta$ and $r = 3\beta - 2/\gamma$. 
They are corresponding to PSs. 
One is a stable PS and the other is an unstable one. 

In the present case of $\gamma < 0$, 
the stable outer PS is located at 
\begin{align}
r_m = 3\beta - \frac{2}{\gamma} , 
\label{rm-Weyl}
\end{align} 
because 
\begin{align}
 D_m' 
=& 
- \frac{2}{r_m^2 g(r_m)} 
\notag\\
=& 
 - \frac{2 b_m^2}{r_m^4} 
 \notag\\
 <& 
 0 . 
\label{Dm1-Weyl}
\end{align}
The other root $r = 3\beta$ is the radius of 
the unstable inner PS. 

Note that $r_m$ is larger than $3\beta$ because of $\gamma < 0$. 
See Figure \ref{fig-V} 
for the effective potential $V(r)$ in the Weyl gravity model 
with the stable outer PS.

There is a constraint on $\kappa$ as 
\cite{footnote-Horne}
\begin{align}
\kappa < - \frac{\gamma^2 (1 - \beta\gamma)}{(2 - 3 \beta\gamma)^2} . 
\label{kappa}
\end{align}

Here, $b_m$ denotes the critical impact parameter 
corresponding to the stable PS. 
It is obtained as 
\begin{align}
b_m = \frac{2 - 3 \beta\gamma}
{\sqrt{-\kappa (2 - 3 \beta\gamma)^2 - \gamma^2(1 - \beta\gamma)}} , 
\label{bm-Weyl}
\end{align}
where the inside of the square root is always positive 
when $\kappa$ satisfies Eq. (\ref{kappa}).

\begin{figure}
\includegraphics[width=8.6cm]{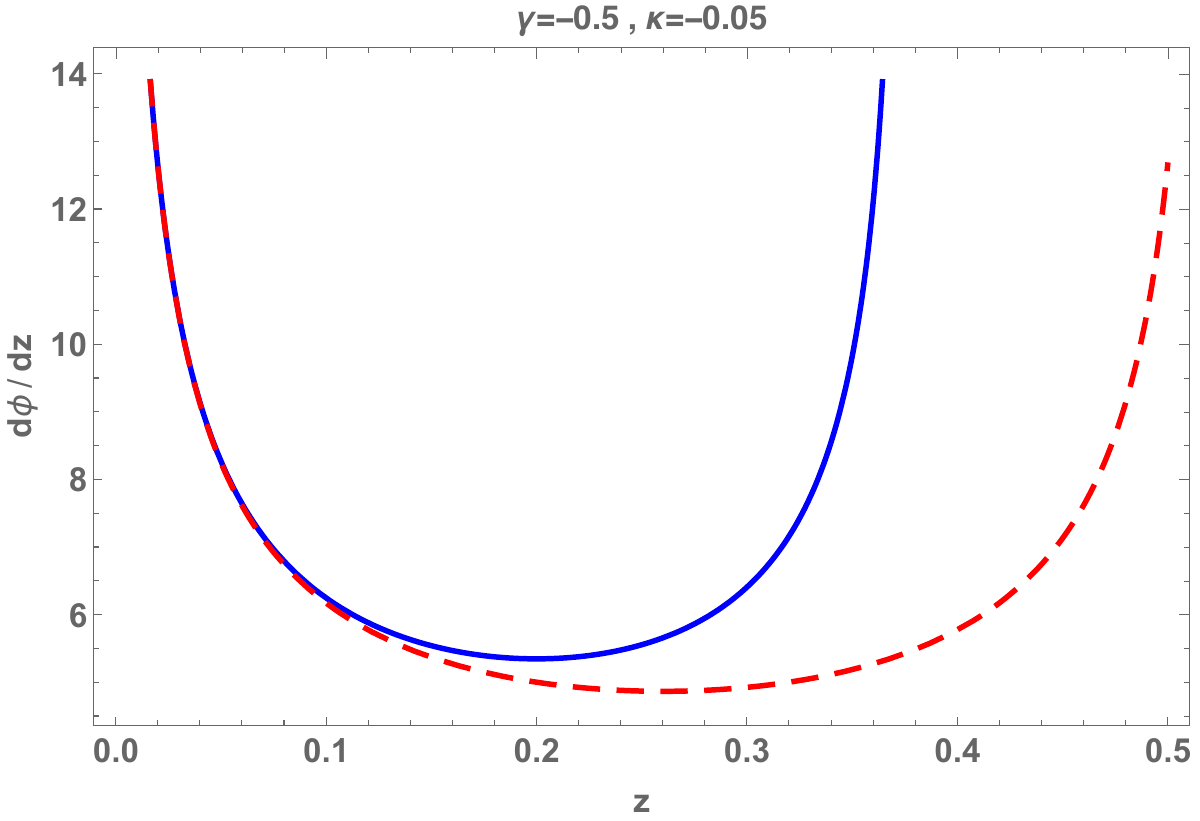}
\includegraphics[width=8.6cm]{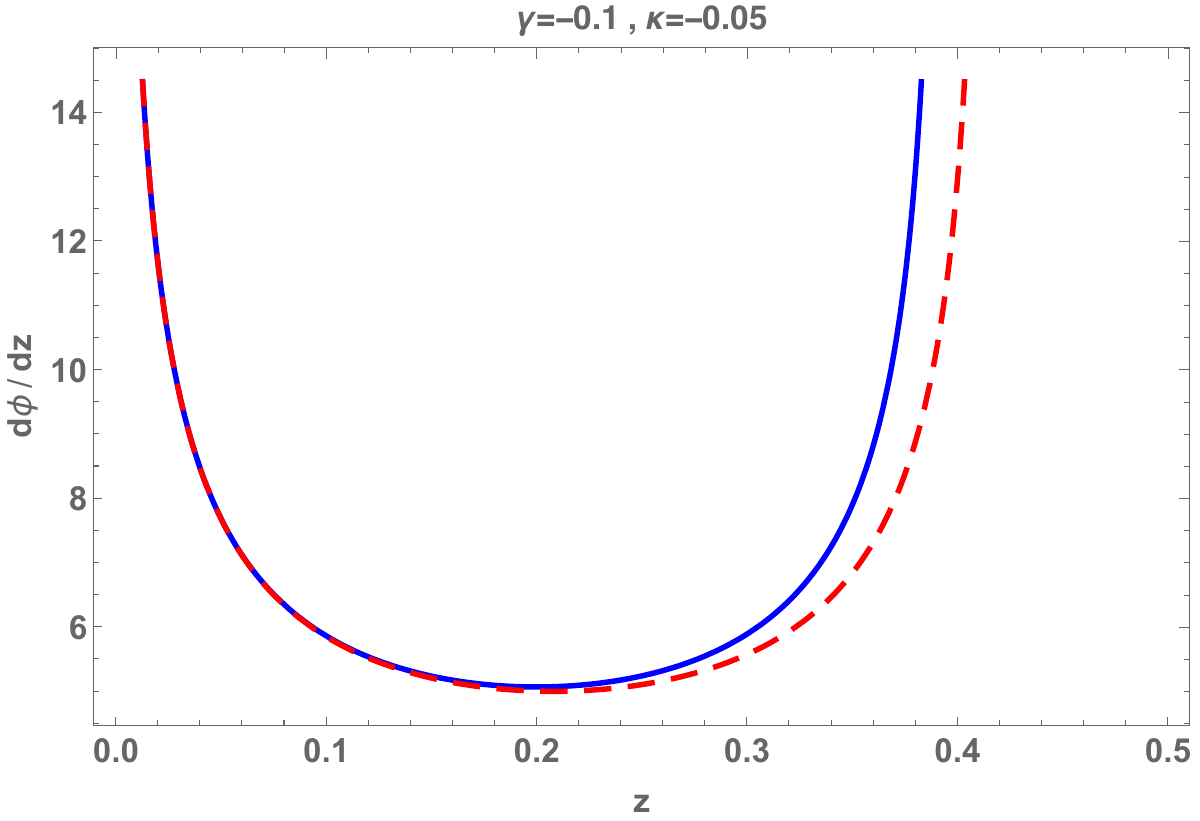}
\caption{
The integrands of Eqs. (\ref{IF-2}) and (\ref{IFD}). 
The solid blue (in clor) curve denotes $f$ ($= d\phi/dz$) 
in the integral $I_F$, 
while the dashed red (in color) curve denotes $f_D$ in $I_{FD}$. 
The horizontal axis means $z$. 
We assume $\kappa = -0.05$ in the unit of $\beta =1$ 
that is roughly corresponding to the conventional unit as mass = 1 
in the Schwarzschild or Kottler spacetime. 
The top and bottom panels assume 
$\gamma = -0.5$ and $\gamma = -0.1$, respectively, 
each of which leads to $r_m = 7.0$ and $b_m = 7.18185$, 
and 
$r_m = 23$ and $b_m = 4.56815$, respectively. 
In both cases, the solid and dashed curves are 
overlapped better for smaller $z$. 
For instance, the difference between them is 
about 10 percents at $z \sim 0.2$. 
A significant deviation at large $z$ is due to 
a departure from the quadratic approximation of $H(z)$ in $z$.
} 
\label{fig-f}
\end{figure}

From Eqs. (\ref{IFD-approx}) and (\ref{IFR-approx2}), 
we obtain 
\begin{align}
&I_{FD}(z_S, z_R, b) 
  \notag\\
 &\simeq \pi - \sum_{i=S,R} \arcsin \left\{ 1 - \frac{b_m}{\sqrt{2} r_m} 
 \left( 1 - \frac{b}{b_m} \right)^{-1/2} z_i \right\} ,
\label{IFD-Weyl}
\end{align}
\begin{align}
&I_{FR}(z_S, z_R, b) 
 \notag\\
&\simeq - \beta \gamma \left( \frac{b_m}{2 \sqrt{2} r_m} \right)^{3/2} \left( 1 - \frac{b}{b_m} \right)^{-3/4} \sum_{i = S,R} z_i^{5/2} ,
\label{IFR-Weyl}
\end{align}
where straightforward calculations at $O(z^{5/2})$ are done   
in Eq. (\ref{IFR-approx2}). 

When $z_T \equiv z_R = z_S \ll (1 - b/b_m)^{1/2}$, 
Eq. (\ref{IFD-Weyl}) provides an approximate expression 
of the deflection angle as 
\begin{align}
\Delta\phi 
=&  
I_F 
\notag\\
= 
&
2 \sqrt{\frac{\sqrt{2} b_m}{r_m} 
\left( 1 - \frac{b}{b_m} \right)^{-1/2} z_T } 
\notag\\
&
+ O\left( \frac{b_m}{r_m} \left( 1 - \frac{b}{b_m} \right)^{-1/2} z_T \right). 
\label{Delta-phi}
\end{align}
where we use 
$\arcsin(1 - \varepsilon) = 
\pi/2 - \sqrt{2 \varepsilon} + O(\varepsilon)$ 
for $\varepsilon < 1$.

\begin{figure}
\includegraphics[width=8.6cm]{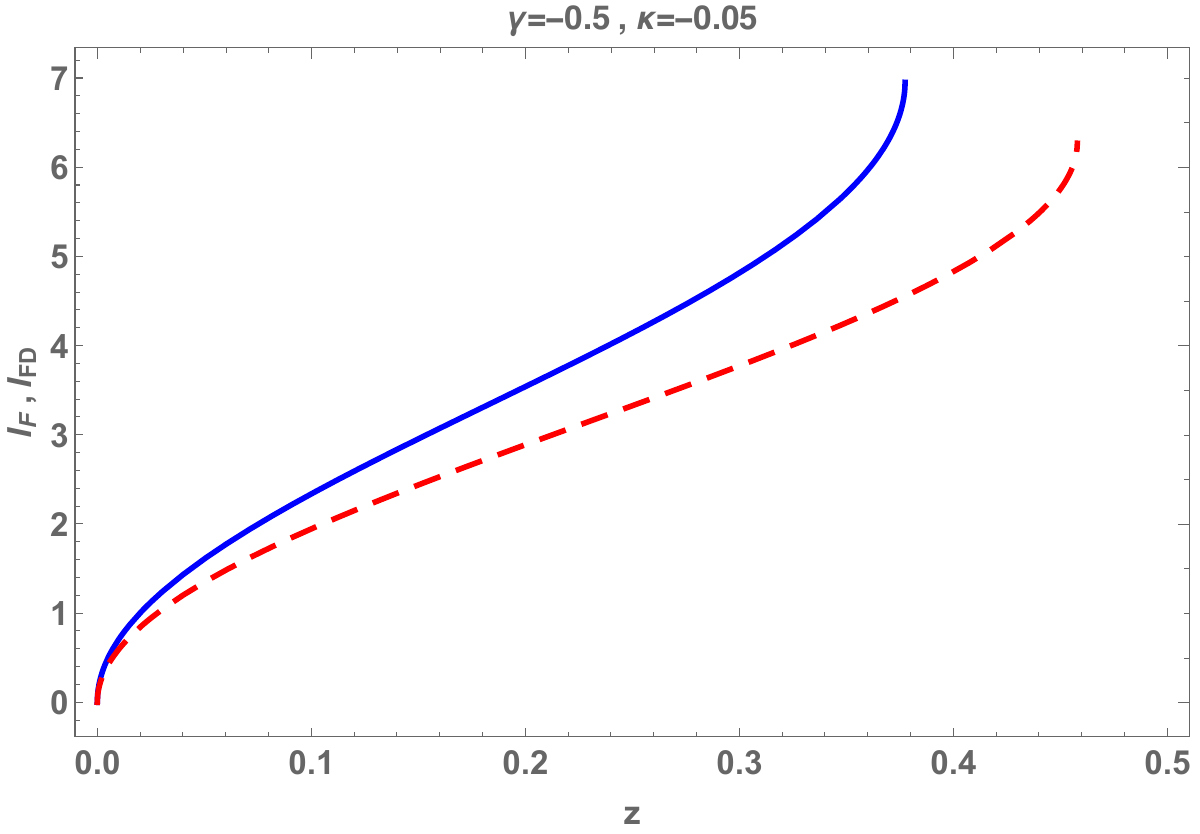}
\includegraphics[width=8.6cm]{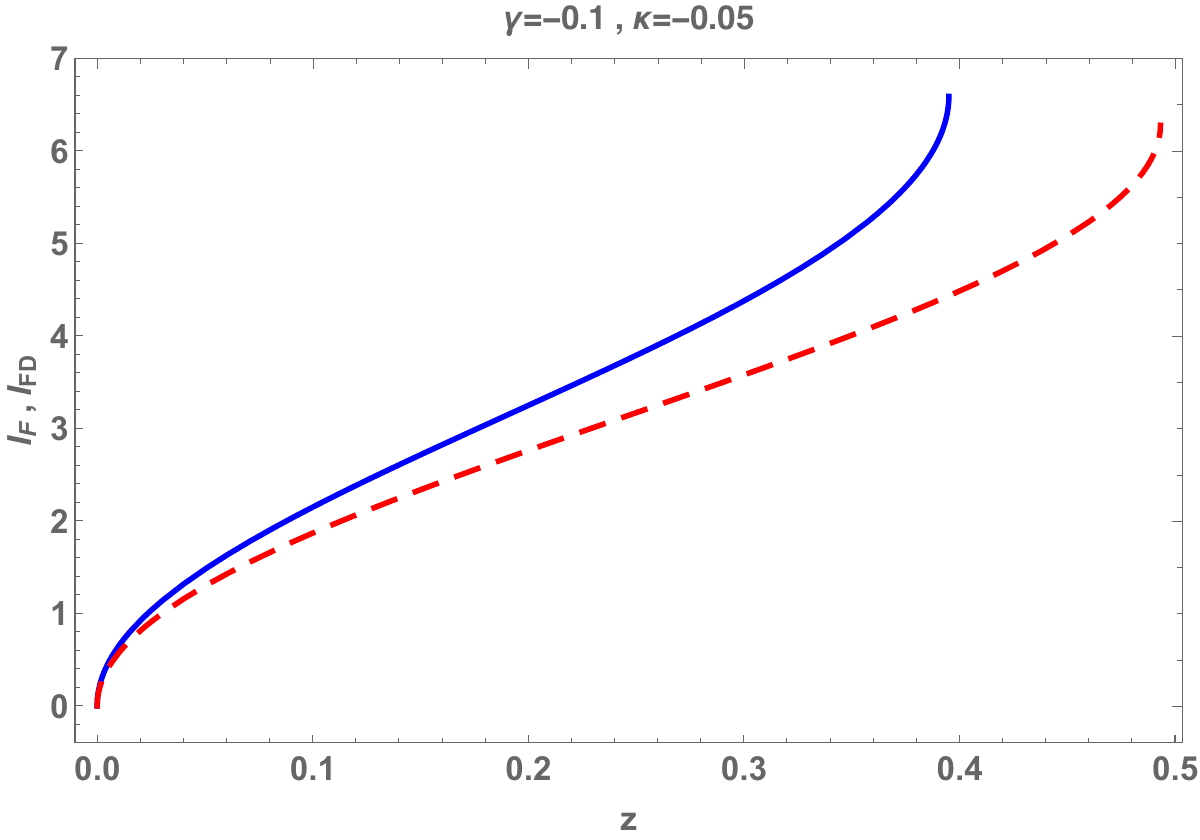}
\caption{
The total angle integrals $I_F$ and $I_{FD}$. 
The solid blue (in color) curve denotes the present approximation 
of $I_{FD}$ by Eq. (\ref{IFD-Weyl}) 
and the dashed red (in color) curve denotes numerical calculations 
of $I_{F}$ in Eq. (\ref{IF-2}), 
where we assume 
the same values for $\kappa$ and $\gamma$ in Figure \ref{fig-f}. 
For the simplicity, we choose $z \equiv z_R = z_S$, 
which is denoted by the horizontal axis. 
For $z \ll1$, the two curves are very close to each other, 
because the quadratic approximation works well 
especially near $z \sim 0$. 
The two curves are close to each other 
especially for smaller $z$. 
For instance, the difference between them is 
roughly 20 percents at $z \sim 0.2$. 
A significant deviation at large $z$ reflects 
a departure from the quadratic approximation of $H(z)$ in $z$.
} 
\label{fig-I}
\end{figure}

It is natural that $I_{FR}$ is much smaller than $I_{FD}$, 
as discussed in Section III. 
Eq. (\ref{IFD-Weyl}) shows the mild deflection in terms of the arcsine function. 
See Figure \ref{fig-I} for a comparison of Eq. (\ref{IFD-Weyl}) 
and numerical calculations for $I_{FD}$. 
As shown in Figure \ref{fig-f}, 
$f$ are dependent on rather strongly on $\gamma$. 
The difference between the numerical $f$ and approximate one 
becomes significant as $z$ is larger. 
On the other hand, the closest approach and its vicinity make 
a dominant contribution to the total angle integral $I_F$. 
In Figure \ref{fig-I}, therefore, 
$I_F$ shows much weaker dependence 
on $\gamma$ and $z$.

Finally, we discuss the gap size. 
By using Eq. (\ref{Dm1-Weyl}) for Eq. (\ref{Deltab}), 
we obtain 
\begin{align}
\Delta b 
= 
\frac{b_m^3 z_L^2}{8 r_m^2} . 
\label{Deltab-Weyl1}
\end{align}
For $b_m \sim r_m$, furthermore, 
it becomes simply 
\begin{align}
\Delta b 
\approx 
\frac18 b_m z_L^2 . 
\label{Deltab-Weyl2}
\end{align}

In a SSS spacetime, 
the photon sphere was thought 
to be always an edge such as 
the inner boundary of a black hole shadow. 
However,  Eqs. (\ref{Deltab-Weyl1}) and (\ref{Deltab-Weyl2}) 
provide a counterexample, 
when the PS is stable.

\section{Conclusion}
We demonstrated that the location of a stable PS  
in a compact object is not always an edge 
such as the inner boundary of a black hole shadow. 
We showed also that
a SSS spacetime 
cannot be 
asymptotically flat, 
when the SOPS exists in the spacetime. 

We proved a proposition that 
the closest approach of a photon 
is prohibited in the immediate vicinity of the stable PS 
when the photon is emitted from a source 
(or reaches a receiver) distant from a lens object. 
We discussed the gap size. 
Because of the existence of the gap, 
the mild deflection is caused for a photon traveling  
around the stable PS 
which exists in a class of  SSS spacetimes. 

Finally, we used a class of SSS solutions in Weyl gravity 
in order to exemplify the mild deflection near the stable outer PS. 
It would be interesting to examine whether 
a light ray is mildly deflected around a stable photon surface 
in a less symmetric or dynamical spacetime. 
It is left for future.

\begin{acknowledgments}
We would like to thank Mareki Honma for the conversations 
on the EHT method and technology. 
We wish to thank Keita Takizawa, Naoki Tsukamoto, Chul-Moon Yoo, 
Kenichi Nakao, Tomohiro Harada and Masaya Amo 
for the useful discussions. 
We thank
Yuuiti Sendouda, Ryuichi Takahashi, Masumi Kasai, 
Kaisei Takahashi, and Yudai Tazawa 
for the useful conversations. 
This work was supported 
in part by Japan Society for the Promotion of Science (JSPS) 
Grant-in-Aid for Scientific Research, 
No. 20K03963 (H.A.),  
in part by Ministry of Education, Culture, Sports, Science, and Technology,  
No. 17H06359 (H.A.).  
\end{acknowledgments}


\begin{thebibliography}{99}
\bibitem{footnote-expansion}
There is a prohibited region around the stable PS as shown in Section IIID. 
Because the existence of this gap region, 
the functions $H$ and $f$ are 
not differentiable at $r = r_m$. 
As a result, $H$ and $f$ cannot be Taylor-expanded around $r_m$. 
In stead of Taylor expansions, therefore, 
we look for asymptotic expansions of $I_F$, $I_{FD}$ and $I_{FR}$ 
through $f$, $f_D$ and $f_R$.
\bibitem{footnote-Horne}
Turner and Horne realized that a bound orbit of a photon 
is possible in this parameter region for the SSS solution 
in Weyl gravity \cite{TH}.
\end{thebibliography}
\end{document}